# Percolation threshold in cluster-assembled nanogranular palladium films and related strategies for hydrogen detection

A. Debot[1], D. Balakrishnan[1], A.M. Philippe[1] and E. Barborini[1]

[1]Luxembourg Institute of Science and Technology (LIST), 41 rue du Brill, 4422 Belvaux

Corresponding author: emanuele.barborini@list.lu

**Abstract**

Electrical transport in cluster-assembled nanogranular palladium films was investigated in-situ during growth by Supersonic Cluster Beam Deposition. The percolation onset is observed at a thickness of 1.5 nm, while the percolation completion at 2.5 nm. Films at the percolation thickness (2 nm) and beyond the percolation end (5 nm) were subsequently exposed ex-situ to 0.25% hydrogen in air to investigate how hydrogen-induced structural changes in Pd clusters modulate electrical transport and can be exploited as mechanism for selective hydrogen detection. Implantation of palladium clusters in soft polymer styrene–ethylene–butylene–styrene was also explored to promote clusters separation and enhance quantum tunnelling phenomena within electrical transport. The observed electrical transport pattern features stable baselines in air, suggesting material stability, fast response time ($T_{50\%}$ < 10 s), and the coexistence of reversible and irreversible components in conduction changes induced by hydrogen exposure. The irreversible component is tentatively attributed to hydrogen-promoted partial coalescence of Pd clusters. The behaviour of the polymer implanted samples suggests an alternative gas detection mechanism where hydrogen-induced polymer swelling dominates over palladium structural modifications. These results disclose novel possible transducing routes for miniaturized hydrogen detection and multimodal sensing.

Hydrogen represents one of the main actors in worldwide energy transition efforts as promising solution for energy storage and transport, to be paired with renewable energy sources. The ubiquitous introduction of hydrogen infrastructures and related safety concerns[1] pushes for the development of monitoring networks based on miniaturized, low power consumption, fast and -above all- selective detectors. The US Department of Energy summarized the main requirements for such devices: concentration range 0.1 to 10 vol% in ambient air; response time < 1 s; operating temperature range -30 °C to 80 °C; lifetime > 10 years; cost of few tens of USD per unit[2]. No off-the-shelf solutions meeting all these requirements currently exist.

It is well established that palladium uniquely promotes the dissociation of hydrogen molecules at its surface followed by interstitial diffusion of hydrogen atoms within the metal lattice[3,4]. This is a highly selective process towards hydrogen leading to the formation of a diluted solid solution of H in Pd (α phase) first, then to the nucleation and growth of the PdH hydride phase (β phase), as the concentration of H increases[5-7]. Since the lattice constant of the β phase is 3.6% larger than the metallic phase (4.03 vs. 3.89 Å, respectively), such structural change can be exploited as selective transduction mechanism for hydrogen detection. A comprehensive survey of hydrogen detection strategies based on this transducing mechanism is provided in recent review articles[8-10]. Among the many, it is worth mentioning the pioneering work of Favier et al. proposing the use of Pd nanowires arrays, where the exposure to hydrogen induces the closing of nanoscopic gaps in the wires upon dilation of palladium grains absorbing hydrogen, which results in the overall decreasing of the resistance of the nanowires array[11]. Response time < 0.1 s to concentrations in the range 2% to 10% was reported.

The use of nanoscale structures is a common trait that introduces benefits in terms of faster response-recovery dynamics and of reduction, or even absence, of the hysteresis affecting property changes upon hydrogen exposure and removal in bulk Pd[12,13]. Except in few cases[11,14], a detailed investigation of the electrical transport in palladium-based nanostructures and the relationship to their morphological features during the growth and the exposure to hydrogen is missing. In this respect, a deeper study of the various phenomena characterizing the electrical transport in cluster-assembled nanogranular Pd films, namely: quantum tunnelling, percolation, resistive switching, may widen the available options for transducing mechanisms.

In this communication, we report about the electrical transport behaviour of cluster-assembled nanogranular Pd films upon hydrogen exposure. Samples were prepared by Supersonic Cluster Beam Deposition (SCBD)[15] (**Fig. 1a**). Special focus was devoted to the characterization of the percolation threshold, in order to identify the thickness range from percolation onset to its end. Samples having thickness within this range and samples well beyond the end of the percolation regime were afterwards prepared aiming to highlight the different impact of hydrogen-induced microstructural rearrangements on electrical transport for these two classes of nanogranular system. The underlying hypothesis is that hydrogen absorption in Pd clusters is promoted by the nanogranular nature of the film. In addition, the dimensions of the constituent particles in the nanometre range are expected to promote fast dynamics in absorption and desorption. A sample version where Pd clusters are implanted into a soft polymer is also explored with respect to hydrogen-induced effect on electrical transport.

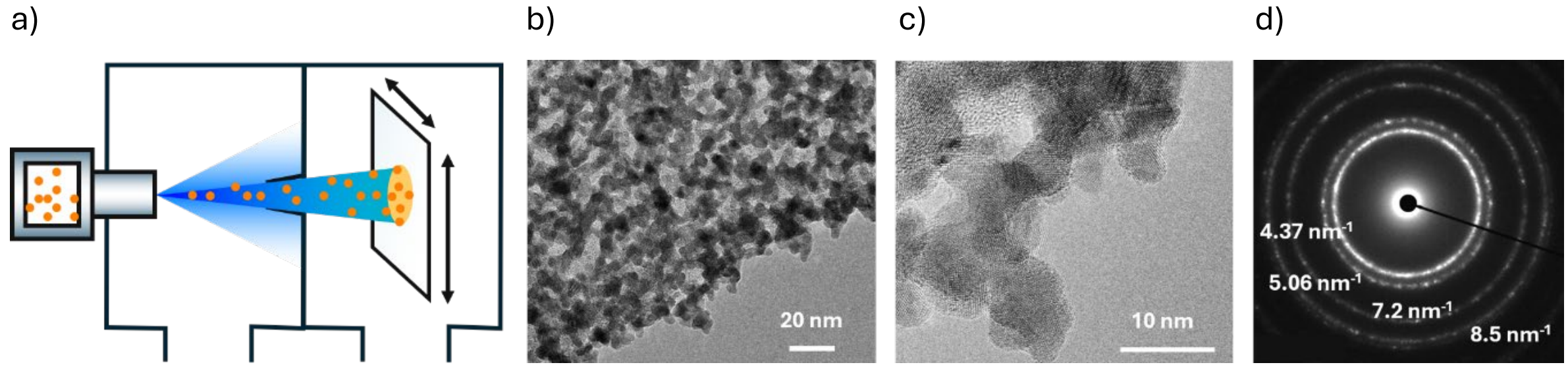

**Figure 1. a)** Scheme of Supersonic Cluster Beam Deposition system, where arrows indicate substrate movements implementing rastering protocols. **b-c)** TEM images of cluster-assembled Pd nanogranular film. **d)** Electron diffraction pattern of the same, matching cubic metallic Pd.

**Fig. 1b-c-d** shows Transmission Electron Microscopy (TEM) images and Selected Area Electron Diffraction (SAED) pattern of the films under investigation. TEM images highlight the nanogranular structure of the films, resulting from the soft assembling of Pd nanoparticles with dimensions <5 nm, while SAED pattern stems from the (111), (200), (220), and (311) planes of face-centered cubic metallic palladium, matching the reference ICDD Powder Diffraction File (Card No. 00-005-0681)[16], as reported in the Table below. The relationship between d and the miller indices is $d = \frac{a}{\sqrt{h^2 + k^2 + l^2}}$, where a = 3.8889 Å of metallic cubic Pd, while the I(%) column represents the relative intensity of the electron diffraction peaks in the reference PDF card.

| Electron diffraction reference (PDF 00-005-0681) | | | | | | This work |
|---|---|---|---|---|---|---|
| d (Å) | I (%) | h | k | l | 1/d ($nm^{-1}$) | SEAD 1/d ($nm^{-1}$) |
| 2.246 | 100 | 1 | 1 | 1 | 4.45 | 4.37 |
| 1.945 | 42 | 2 | 0 | 0 | 5.14 | 5.06 |
| 1.376 | 25 | 2 | 2 | 0 | 7.27 | 7.2 |
| 1.173 | 24 | 3 | 1 | 1 | 8.53 | 8.5 |

For the investigation in-situ of the percolation, we took advantage of a motorized sample holder onto which a pair of interdigitated electrodes (IDEs) made of gold fingers with 100 µm spacing deposited on silica substrate were positioned beside a quartz crystal microbalance (QCM) head. Details of the experimental setup can be found in J.E. Martinez Medina et al.[17]. IDEs and QCM were horizontally rastered in front of the cluster beam for them to collect exactly the same amount of Pd clusters, enabling current-thickness correlation at the end of each rastering scan. Due to the working principle of the QCM, the measured thickness here must be intended as the "bulk-equivalent thickness" of the cluster-assembled Pd film, i.e. the thickness of a compact Pd film having mass per unit area corresponding to that of the nanogranular film. The voltage applied to IDEs during in-situ conduction measurement was set to 50 mV. The SCBD system featured a base pressure (cluster source off) of about $2\times10^{-8}$ mbar. **Fig. 2** shows the results of percolation threshold characterization. The onset occurs at about 1.5 nm while the termination at about 2.5 nm, as clearly visible in the semi-logarithmic inset graph, zooming on 1-4 nm thickness range.

Based on percolation results, two sets of samples were prepared: one at 2 nm, in the middle of percolation; one at 5 nm, well beyond the percolation. Samples were obtained by depositing Pd clusters onto IDEs identical to the ones for in-situ measurements. A third set of samples was prepared, where the amount of Pd nanoparticles corresponding to 5 nm was incorporated into styrene–ethylene–butylene–styrene (SEBS) block co-polymer, taking advantage of the implantation phenomenon occurring when supersonic cluster beams are directed towards soft polymers[15,18,19]. Implantation in SEBS would promote the formation of a polymer-nanoparticles composite layer, where Pd clusters are not necessarily in contact with each other, and the presence of a very thin polymeric layer between at least a part of Pd clusters is expected to emphasize the quantum tunnelling contribution to the overall electrical transport. As a thermoplastic elastomer, SEBS was selected because gas transport in this material occurs via the solution–diffusion mechanism, enabling hydrogen diffusion through the polymer[20,21]. After implantation, gold metallic pads were evaporated on polymer surface exposed to the Pd cluster beam for electrical readout.

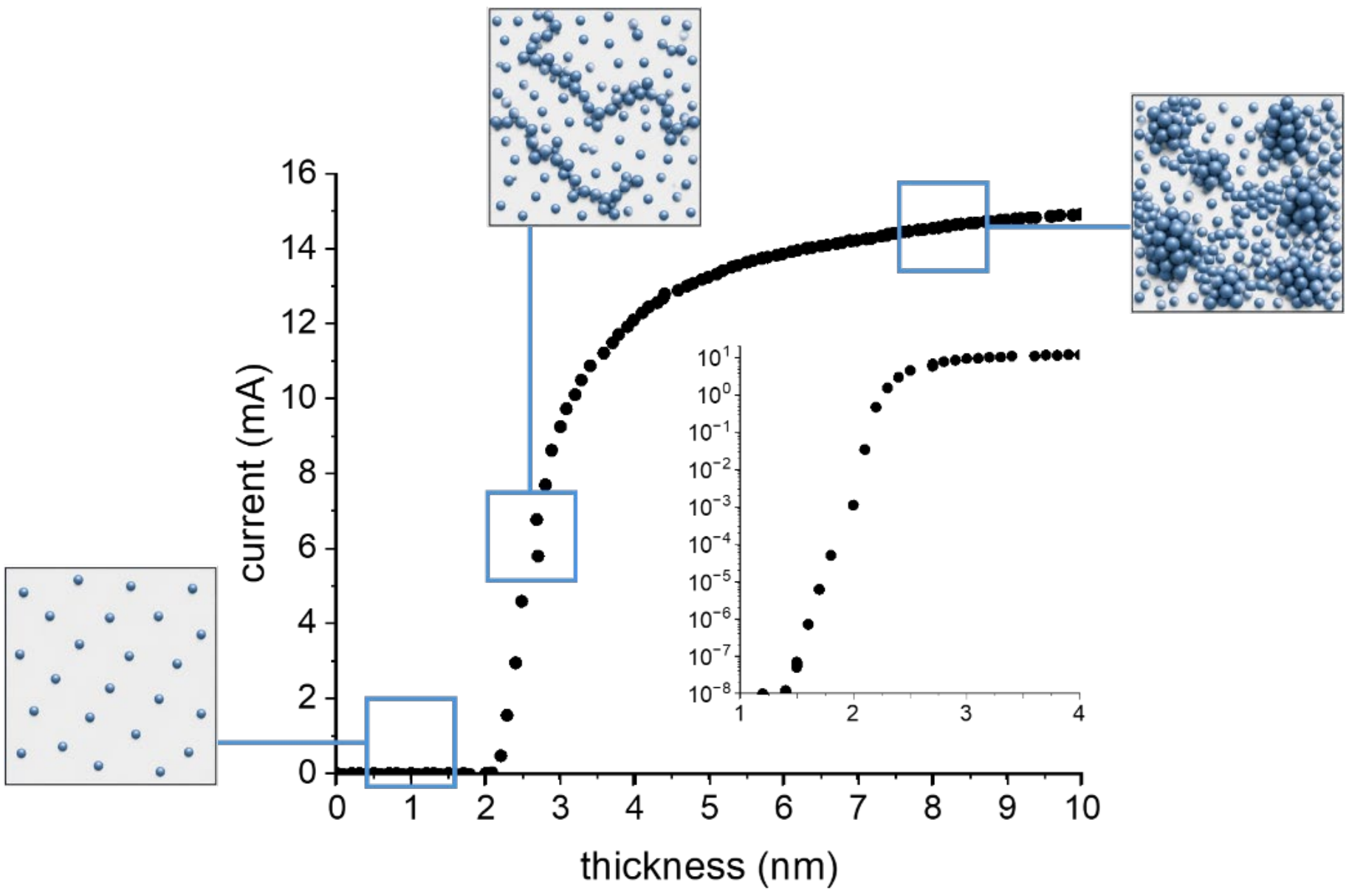

**Figure 2.** In-situ current-thickness correlation during nanogranular Pd film growth. The percolation phenomenon occurs between 1.5 and 2.5 nm, as clearly visible in semi-log inset graph zooming on 1-4 nm thickness range. Images from left to right schematically depict the three main coverage-conduction stages: no conduction, percolation, conduction through the 3D cluster-assembled structure.

Hydrogen sensing measurements were carried out in a dedicated test chamber, where Pd samples were alternately exposed to 0.25% $H_2$ in air (volume), and to air only. The volumetric concentration of 0.25% is a reasonable value for leakage detection and well below hydrogen lower explosive limit (LEL) in air, i.e. 4%. Exposure time and purging time were both set at 2 h: long enough to appreciate very slow dynamics in response and recovery. Four exposure-purging sequences were executed following an initial purging time in pure air with 4 h duration. Since temperature variations are expected to impact hydrogen absorption, samples were kept at 50°C during the measurement period to avoid effects of night-day ambient temperature fluctuations. Voltage applied to IDEs for current readout was 50 mV.

In samples at percolation (2 nm), $H_2$ exposure causes a sudden and rapid increase in conduction (~2%), as shown in top graph of **Fig. 3a**. The time to half-maximum response ($T_{50\%}$) is < 4 s (top graph **Fig. 3b**). Subsequently, the trend reverses, and a steady decrease in electrical current is observed during continued exposure. The flat baseline following exposures suggests sample stability in air, while the absence of recovery indicates that irreversible mechanisms dominate the transport behaviour. At percolation, quantum tunnelling is expected to contribute to electrical transport mechanisms. We propose that initial Pd lattice expansion upon hydrogen absorption narrows the tunnelling gaps and creates new points of contact, both leading to the observed fast-dynamics increase in conduction[14]. However, these microstructural rearrangements may promote coalescence, progressively disrupting -with slower

dynamics- some of the conduction pathways that sustain percolation. Once hydrogen is removed, these disconnected pathways cannot spontaneously reform, which may account for the progressive decrease of conduction with no baseline recovery.

In samples with thickness well beyond percolation completion (5 nm), electrical transport is expected to occur through the 3D metallic network characterizing the nanogranular film and formed by the soft assembling of Pd clusters. The effect of hydrogen exposure on conduction is small (< 0.5 %), although clearly visible (mid graph of **Fig. 3a**). Upon $H_2$ exposure, conduction exhibits a sudden and rapid increase ($T_{50\%}$ ~ 10 s, mid graph **Fig. 3b**), followed by a slower, steady rise during the exposure period. The stability of the baseline in pure air confirms that the observed changes are induced solely by hydrogen exposure. In this regard, lattice expansion of Pd nanoparticles is assumed to promote particle approach and the formation of additional contact points, leading to the fast-dynamics increase of conduction in the presence of $H_2$. Hydrogen removal restores larger interparticle gaps, causing the observed fast-dynamics decrease in conduction (reversible component). Coalescence phenomena occurring during hydrogen exposure may explain the non-complete recovery (irreversible component) and related conduction drift.

In polymer implanted samples, the formation of the metallic 3D network dominating electrical transport in the previous samples is inhibited, at least partially. A significant fraction of Pd nanoparticles is expected to be surrounded by a very thin polymer layer, amplifying quantum tunnelling contribution to the conduction mechanism. At the onset of $H_2$ exposure, a sudden and rapid decrease in conduction is observed (-40%), followed by a gradual increase (bottom graph of **Fig. 3a**). The time to half-maximum response $T_{50\%}$ is ~ 8 s (bottom graph **Fig. 3b**). As in previous samples, the stability of the baseline in pure air confirms that the observed changes originate from hydrogen exposure. The initial decrease in conduction is attributed to the enlargement of tunnelling gaps due to polymer swelling upon hydrogen absorption, which dominates over Pd lattice expansion. A fraction of nanoparticles not fully separated by the polymer may still undergo (partial) coalescence, favoured by Pd lattice expansion and microstructural rearrangements. This mechanism can account for both the gradual increase in conduction during hydrogen exposure and the recovery to a higher baseline after hydrogen removal. Polymer swelling is proposed as the reversible component of the electrical conduction changes upon hydrogen exposure, while hydrogen-promoted partial coalescence of Pd clusters as the irreversible component.

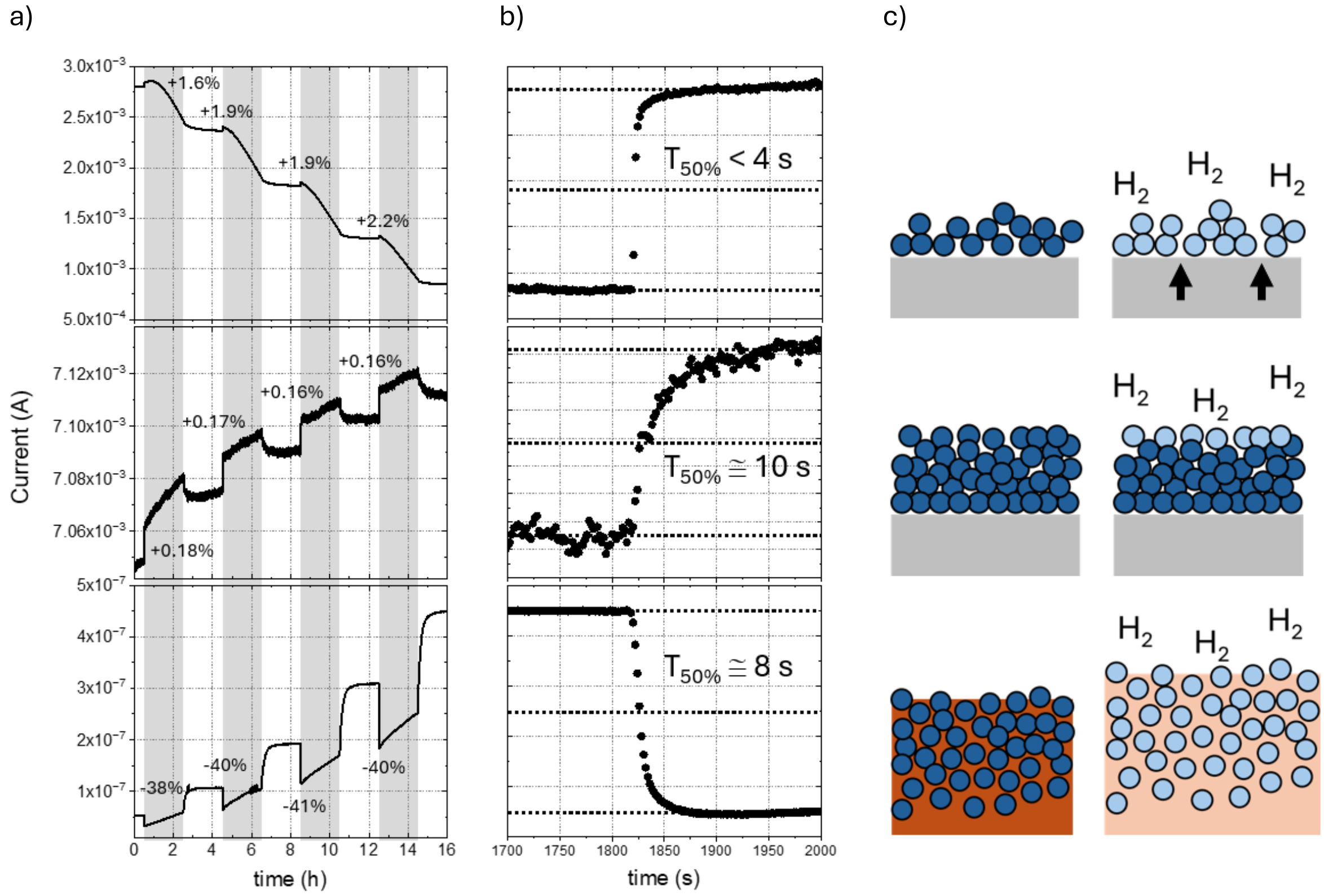

**Figure 3**. **a)** Hydrogen detection by the three nanogranular Pd structures described in the text: percolating one (top); 3D metallic network (middle); polymer-nanoparticles composite layer (bottom). The electrical current through the films was measured while alternating exposures to 0.25% $H_2$ in air and pure air, with each exposure lasting 2 h. Grey bars in the left panels indicate hydrogen exposure periods. The percentage change in current at the onset of each exposure is reported. **b)** Magnified views of the first $H_2$ exposure onsets, highlighting the fast responses with $T_{50\%}$ of 4, 10, and 8 s respectively. **c)** Schematic cross-sectional views of the films illustrating the mechanisms proposed to explain the observed electrical transport behaviour: the grey rectangle represents the dielectric substrate ($SiO_2$); dark blue circles represent Pd clusters not interacting with hydrogen; light blue circles represent Pd clusters interacting with $H_2$; the dark brown rectangle denotes the polymer embedding the Pd nanoparticles; and the light brown rectangle represents the polymer swollen due to hydrogen absorption. The black arrows in the scheme of the material at percolation (top) highlight the points where conduction path gets disrupted.

Hydrogen-induced modifications of the microstructure of nanogranular Pd films obtained by cluster beam deposition are translated in electrical transport modulation, which has been demonstrated to enable hydrogen detection at concentrations of applicative interest. According to the amount of Pd clusters deposited with respect to percolation threshold, various phenomena affecting the electrical transport are observed. These include a conduction increase with very fast dynamics upon hydrogen exposure, tentatively attributed to the quantum tunnelling component of conduction mechanisms and to the creation of new points of contacts between particles. The conduction increase is partially reversible at hydrogen removal in films with thickness beyond percolation. In films with thickness at percolation the conduction increase represents a fast-dynamic initial event ($T_{50\%} < 4$ s), which is subsequently reversed while hydrogen exposure continues.

Samples obtained by cluster beam implantation in soft polymer SEBS respond to hydrogen with an electrical transport pattern suggesting polymer swelling dominates the modulation of inter-particle gaps and related impact on the quantum tunnelling component of the conduction. In this respect, Pd specificity towards hydrogen is not exploited as detection mechanism, which is expected to be non-selective. This however suggests transferring to the polymer the selectivity function, by fine tuning the solution–diffusion mechanism and related swelling through proper formulation. Such chemoresponsive polymers will then be implanted with metallic nanoparticles, which are not required to be of any specific kind. Hygroscopic polymers, such as for example polyvinyl alcohol (PVA) or polyacrylic acid (PAA) among the many, may enable humidity sensing capabilities based on the described mechanism.

Interestingly, all sample configurations studied show stable baseline conduction in pure air, suggesting that coalescence phenomena are exclusively hydrogen-induced. Partial coalescence of Pd clusters is proposed to underlie all the irreversible changes in conduction, specifically causing the disruption of conduction paths in films at percolation. In this respect, a further research direction to explore would regard samples in pre-percolation conditions (isolated clusters): here quantum tunnelling is expected to dominate the conduction mechanism, hydrogen absorption would modulate the tunnelling gaps ensuring a detection mechanism, but the absence of any established conduction path would likely prevent coalescence-related irreversible phenomena.

Overall, the presented results suggest cluster-assembled nanogranular Pd films as possible enabling system for selective, miniaturized, low power consumption, hydrogen detectors enabling capillary monitoring of hydrogen infrastructures for energy storage and mobility. Monitoring is expected to improve safety, mitigate citizen concerns, and favour public acceptance. Though renouncing to palladium-related hydrogen specificity, polymer-nanoparticles composite layers at percolation, obtained by cluster beam implantation, may offer a platform for multimodal sensing, including humidity and volatile chemicals sensing, this last achieved through proper polymer formulation. In addition, such nanocomposite layers at percolation are expected to also feature high-sensitivity touch sensors thanks to the exponential link between electrical conduction and particles per unit area, a parameter modulated by mechanical deformations.

**Acknowledgements**

The authors acknowledge financial support from the Luxembourg National Research Fund (FNR), CLASMARTS project (C19/MS/13685974). The authors thank R. Leturcq for support on the use of the gas sensing testbench at LIST.

**Conflict of Interest**

The authors have no conflicts to disclose.